# On the Euler angles for the classical groups, Schwinger approach and Isoscalar factors for SU (3)

**M. Hage-Hassan**
Université Libanaise, Faculté des Sciences Section (1)
Hadath-Beyrouth

## Abstract

We establish recurrences formulas of the order of the classical groups that allow us to find a generalization of Euler's angles for classical groups and the invariant measures of these groups. We find the generating function for the SU(2)$\subset$ SU(3) basis in the Fock-Bargmann space and a new basis of SU(3). This new basis is eigenfunction of the square of kinetic moment in product spaces of spherical harmonics. We generalize the generating function of SU (2) and we find invariant polynomials of SU (3) which are elements of the basis of SU (6). Using the above results we deduce the method of calculation isoscalar factors. We expose this method and we give the generating function for a particular case. Finally we determine the generating function of the elements of the representation matrix of SU(3) and we derive the analytical expression of these elements.

## 1-Introduction

The applications of the SU (n) group theory have occurred in numerous research areas, nuclear physics, high-energy particle theory and experimental nano-scale physics. Several problems remain under investigation: the parameterization of these groups [1-3] the Wigner D-functions [4-5] and the determination of Wigner's 3j coefficients. The Wigner's 3j coefficients are very important for applications and are not completely solved despite the extensive efforts that have been put into this problem by many authors [6-22]. A major obstacle encountered in any approach is the outer multiplicity problem and the complicated calculations. The Kronecker product of two irreducible representations of SU(n), $n \geqq 3$, can be represented as a sum of irreducible representations (IR), a given IR may appear more than once.

In the SU(3) case, the problem has been studied by diverse methods and often without the knowledge of previous works[12].The Van der Waerden approach for SU(2),based on the construction of invariant polynomials, has been followed by several authors [10-12]. Among others Resnikoff who follow the works of Moshinsky et al. [6-9], in Bargmann space [23] for the study of the group SU(3), but the difficulty of the calculations and the multiplicity problem had to permit progress. Another approach is the tensor operator method proposed by Biedenharn and collaborators [13-17]. This method was revisited in the late eighties using the Vector Coherent State (VCS) theory, [18-21]. Although the results seem no simpler than those found earlier.

In this work we start from the order of the classical groups to determine recurrences formulas from which we derive the parameterization or the generalization of Euler's angles for these groups. We prove the connexion of the measure of SU(n) with the measure of product of cylindrical basis of harmonic oscillator or two dimensions Fock-Bargmann spaces. Our approach for SU (n) is based on the above connexion, the construction of the generating function and the propriety $SU(n-1)\subset SU(n)$ to determine the generating function of Wigner 3j coefficients and the invariant polynomials or Van der Waerden invariants.

We have already found the generating function [22] of the basis of SU(3) groups by coupling three basis of two dimensions Bargmann basis, or harmonic oscillator basis, using Schwinger coupling method [28]. Instead the basis of Bargmann we can take the basis of spherical harmonics then we find a new basis eigenfunctions of the square of angular momentum. We generalized the generating function of SU(2) basis by a simple and an empirical way for the determination of Van der Waerden invariants which may be found as elements of the basis of SU(6). We deduce then the good coupling of angular momentum for U (3) patterns [17]. Our unambiguous proof [25] of expressions for 3j symbols was reconsidered, together with a calculation of Wigner's D-matrix elements. This paper is intended to give a revision and a comprehensive review of our approach.

In part 2 we review the Wigner's D-matrix and Gel'fand basis for SU (n). Part 3 is devoted to the generalization of Euler angles to the classical groups. In parts 4-7 we construct the generating functions of SU (3) in Bargmann space, in spherical harmonic basis and the generalization to SU (n). In part 7 we consider the U (3) patterns and the SU (2) coupling. Generating function of invariant polynomials of SU (n) and isoscalar factors of SU (3) are in parts 8 - 9. The part 10 is devoted to the derivation of the Wigner's D-matrix elements for SU(3).

## 2-Review of Wigner's D-matrix and Gel'fand basis for SU(n)

We give a quick revision of Fock-Bargmann space, the properties of classical groups, the basis of Gel'fand and of Wigner's 3j symbols for unitary groups.

### 2.1 The Fock-Bargmann space

All analytic functions $\varphi_n(z)=\dfrac{z^n}{\sqrt{n!}}$ form a basis of Hilbert space, which is known by the Fock-Bargmann space or Analytic Hilbert space.

With
$$\langle\varphi_n|\varphi_m\rangle=\iint\overline{\varphi_n(z)}\varphi_m(z)d\mu(z)=\delta_{m,n}\tag{2.1}$$
$d\mu(z)$ is the cylindrical measure
$$d\mu(z)=e^{-(z\bar{z})}\frac{dzd\bar{z}}{\pi},\; z=x+iy, dzd\bar{z}=dxdy\tag{2.2}$$
for applications we use the formulas:
$$\int e^{x\bar{z}}e^{yz}d\mu(z)=e^{xy}\tag{2.3}$$

$$\langle f' \| g \rangle = \langle f \| zg \rangle, \langle \bar{z}f \| g \rangle = \langle f \| g' \rangle \tag{2.4}$$

f, g are polynomials in z.

## 2.2 The classical groups

The classical groups are the unitary, orthogonal and the symplectic. We make a briefly revisions of the properties of these groups.

1-The special orthogonal group SO($n$): the special orthogonal group SO($n$) is the group of $n\times n$ orthogonal matrices $(A_n^0)$ with unit determinant and dimension, the number of parameters, N(n,0)= $n(n$-1)/2.

2- Special unitary group SU($n$): the special unitary group SU($n$) is the group of $n\times n$ unitary matrices $(A_n^1)$ with unit determinant and dimension N(n,1)= $n^2 - 1$.

3-The symplectic group Sp($n$): The symplectic group Sp($n$) is the group of n x n quaternionic matrices $(A_n^2)$ that preserve the standard hermitian form on $Q^n$ with Dimension N (n,2)=n(2n+1).

## 2.3 Gel'fand Basis for SU(n)

The $n^2$ Weyl infinitesimal generators $E_{ij}$ (i, j=1… n) of the unitary group obey the commutation relations

$$[E_{ij}, E_{kl}] = \delta_{jk} E_{il} - \delta_{il} F_{ki} . \tag{2.5}$$

These generators may be written in terms of creations and destruction of n-dimensional harmonic oscillators as:

$$E_{ij} = \sum_{i=1}^{n} a_i^+ a_j \tag{2.6}$$

The Gel'fand basis for SU (n) is:
$$\left| h_{ij} \right\rangle = \left| \begin{matrix} [h]_n \\ [h]_{n-1} \\ (h)_{n-1} \end{matrix} \right\rangle \tag{2.7}$$

With $$[h]_n = [h_{1n}\ h_{2n}\ h_{3n}\ \dots\ h_{nn}]$$

And $$h_{l,m} \ge h_{l,m-1} \ge h_{l+1,m}$$

The application of the unitary transformation to the basis $\left| \begin{pmatrix} [h]_n \\ (h)_n \end{pmatrix} \right\rangle$ is

$$T_{U_n} \left| \begin{pmatrix} [h]_n \\ (h)_n \end{pmatrix} \right\rangle = \sum_{(h')} D^{[h]_n}_{(h'),(h)} (U_n) \left| \begin{pmatrix} [h]_n \\ (h)_n \end{pmatrix} \right\rangle \tag{2.8}$$

The elements of the matrix of SU(n) are $D^{[h]_n}_{(h'),(h)}(U_n)$

The Gel'fand state for which $h_{rs} = h_{rn}$ ,$1 \le r \le s \le n$, is the state of highest weight [14].

A special result which is immediately available from table techniques [14] is the so-called semi-maximal case:

$$D^{[h]_n}_{\binom{[h]_{n-1}}{(\max)},(\max)}(U) = \frac{1}{\sqrt{N}} \prod_{k=1}^{n-1} (u^{(12..k)}_{(12..k)})^{h_{k,n-1}-h_{k+1,n}} \times \prod_{k=1}^{n-1} (u^{(12..k)}_{(12..k-1,n)})^{h_{k,n}-h_{k,n-1}} \tag{2.9}$$

$u^{(12..k)}_{(12..k)}$ Is the minors constructed from the matrix (Un).

The normalization is:

$$N = \prod_{\substack{i<j\\1}}^{n} \frac{(p_{i,n-1} - p_{j,n})!}{(p_{i,n} - p_{j,n} - 1)!} \prod_{\substack{i<j\\1}}^{n-1} \frac{(p_{i,n} - p_{j,n} - 1)!}{(p_{i,n-1} - p_{j,n} - 1)!}$$

$$p_{in} = h_{in} + n - i$$

## 2.4 The conjugate representation and Wigner's symbols

### 2.4.1 The conjugate representation

Define the transformation

$$T_{U_n} \left| \begin{pmatrix} [h]_n \\ (h)_n \end{pmatrix} \right\rangle_c = \sum_{(h')} (D^{[h]_n}_{(h'),(h)}(U_n))^* \left| \begin{pmatrix} [h]_n \\ (h)_n \end{pmatrix} \right\rangle_c \tag{2.10}$$

The conjugate of the basis states is

$$\left| \begin{pmatrix} [h]_n \\ (h)_n \end{pmatrix} \right\rangle_c \quad \text{With} \quad \left( \left| \begin{pmatrix} [h]_n \\ (h)_n \end{pmatrix} \right\rangle_c \right)_c = \left| \begin{pmatrix} [h]_n \\ (h)_n \end{pmatrix} \right\rangle$$

### 2.4.2 The Wigner's symbols

The direct product of two representations may be reduced according to the formula

$$[h^1] \otimes [h^2] = \sum (\rho)[h^3]_\rho \tag{2.11}$$

Where ($\rho$) is the multiplicity or the number of time the representation $[h^3]_\rho$ is contained in $[h^1] \otimes [h^2]$ .

With $$\left| \begin{pmatrix} [h^3] \\ (h^3) \end{pmatrix} \right\rangle_\rho = \sum_{h^1 h^2} \left\langle \begin{matrix} [h^1][h^2] \\ (h^1)(h^2) \end{matrix} \middle| \begin{matrix} [h^3] \\ (h^3) \end{matrix} \right\rangle_\rho \left| \begin{pmatrix} [h^3] \\ (h^3) \end{pmatrix} \right\rangle \left| \begin{pmatrix} [h^3] \\ (h^3) \end{pmatrix} \right\rangle \tag{2.12}$$

The coefficients in this expression are the Clebsh-Gordan coefficients [10-12].

The vector $$\frac{1}{\sqrt{d_{h^3}}} \sum_{h^3} \left| \begin{pmatrix} [h^3] \\ (h^3) \end{pmatrix} \right\rangle_\rho \left| \begin{pmatrix} [h^3] \\ (h^3) \end{pmatrix} \right\rangle_c \tag{2.13}$$

is an invariant by unitary transformation with unity norm in the triple product space.

When we replace it with the above mentioned:

$$H_{(\rho)} = \sum_{h^1 h^2} \frac{1}{\sqrt{d_{h^3}}} \left\langle \begin{matrix} [h^1][h^2] \\ (h^1)(h^2) \end{matrix} \middle| \begin{matrix} [h^3] \\ (h^3) \end{matrix} \right\rangle_{c\rho} \prod_{i=1}^{3} \left| \begin{pmatrix} [h^i]_n \\ (h_i)_n \end{pmatrix} \right\rangle \tag{2.14}$$

The coefficients $$\frac{1}{\sqrt{d_{h^3}}} \left\langle \begin{matrix} [h^1][h^2] \\ (h^1)(h^2) \end{matrix} \middle| \begin{matrix} [h^3] \\ (h^3) \end{matrix} \right\rangle_{c\rho} = \begin{pmatrix} [h^1], & [h^2], & [h^3] \\ (h_1) & (h_2) & (h_3) \end{pmatrix}_\rho$$

Are Wigner's 3j symbols of SU (n), which are defined up to a phase factor. $H_{(\rho)}$ is a invariant polynomial by the transformation U(n) and it is the generalization of the Van der Wearden's invariant of the group SU(2). These invariants has the following properties

$$T_{(U)}^{(1,2,3)} H_{(\rho)} = H_{(\rho)} \tag{2.15}$$

With
$$\left\langle H_{(\rho)} \middle| H_{(\rho')} \right\rangle = \delta_{(\rho),(\rho')} \tag{2.16}$$

These properties mean that the invariant polynomial is function of elementary invariants. These elementary invariants are not independent for n ≥ 3 and as a particular case we choose $H_{(\rho)}$ as subspace of SU(3 (n-1)) which are function of the compatible elementary invariants.

## 3. On the Euler angles for the classical groups

We derive first from two kinds of recurrences relations the parameterization of the classical groups and then the measures of integration on SO (n), SU (n) and the connexion of the measure of unitary groups with the measures of integration in Fock-Bargmann spaces.

### 3.1.1 Generalization of the Euler parameterization of SO(3)

In Quantum mechanic we write the matrix elements of rotation by

$$\langle lm' | R(\psi\theta\varphi | lm\rangle = \langle lm' | e^{-i\psi L_z} e^{-i\theta L_y} e^{-i\varphi L_z} | lm\rangle \tag{3.1}$$

We observe in the above expression that for every rotation in the space $|lm\rangle$ there is a rotation in the dual space $\langle lm'|$. In this interpretation, we can write the finite transformation of classical groups in the form:

$$A_n^m = A_{n-1}^m B_n^m A_{n-1}^m \tag{3.2}$$

With m = 0, 1 and 2 for orthogonal, unitary and symplectic groups.
In the following we derive two kinds of recurrences formulas

### 3.1.2 First recurrences relations for the number of parameters

It's simple to verify the recurrences relations

$$N(n,m) = N(n-1,m) + 2^m n - 1, \; m = 0,1,2. \tag{3.3}$$

So the order of the matrix n has $2^m n - 1$ parameters more than the matrix of order n-1. Since the point $(0,\ldots,0,1)$ is invariant by the group of order n-1 this means that the last column and the last row are the components of the unit vectors of points on the unit sphere $S^{mn-1}$, of the Euclidian space $E_n(K), K = R, C, H = Q$.

$$A_n^m = \begin{pmatrix} \vdots & \cdots & Last \\ \vdots & \cdots & Col. \\ Last & Row & a_{nn} \end{pmatrix} \tag{3.4}$$

### 3.2.3 Second recurrences relations for the number of parameters

It's also simple to verify the recurrences relations

$$N(n,m)=2N(n-1,m)-N(n-2,m)+2^{m},\ m=0,1,2. \quad (3.5)$$

In the expression

$$A_n^m = A_{n-1}^m B_n^m A_{n-1}^m \quad (3.6)$$

It is quite evident that the parameters of left and right are different.

But

$$A_n^m = A_n^m = A_{n-2}^m B_{n-1}^m A_{n-2}^m B_n^m A_{n-2}^m B_{n-1}^m A_{n-2}^m$$
$$= A_{n-2}^m B_{n-1}^m [A_{n-2}^m B_n^m A_{n-2}^m] B_{n-1}^m A_{n-2}^m$$

We choose

$$[B_n^m, A_{n-2}^m]=0 \quad (3.7)$$

Then number of parameters of $A_n^m$ becomes $2N(n-1,m)-N(n-2,m)+2^m$ and the number of parameters of $B_n^m$ is $2^m$. Therefore we find the same result of the recurrence relation (3.5).

Therefore we write $A_n^m = A_n^m = A_{n-2}^m B_{n-1}^m B_n^m A_{n-2}^m B_{n-1}^m A_{n-2}^m$ (3.8)

To find $A_n^m$ we must choose the parameters such that the last line, or the last column, are the components of the vector $\vec{r}=(x_1,x_2,...,x_n)$, $\vec{r}\cdot\vec{r}=1$ and $[B_n^m, A_{n-2}^m]=0$. In this case the range of parameters is imposed by the range of the variation of the angles of the vector $\vec{r}$.

It is important to note that every parameterization components of the vector $\vec{r}$ corresponds to a parameterization of classical groups and therefore the parameterization is not unique.

### 3.3 Parameterization of SO(n)

In this case $A_n^m = A_{n-1}^m B_n^m A_{n-1}^m$, m = 0 the matrix $B_n^0$ is function of one variable and The expression $[B_n^0, A_{n-2}^0]=0$ means that $B_n^0$ leave invariant $A_{n-2}^0$.

Then we write

$$B_n^0=\begin{pmatrix} I_{n-2} & 0 & 0 \\ 0 & \cos\theta_{n-1}^{n-1} & \sin\theta_{n-1}^{n-1} \\ 0 & -\sin\theta_{n-1}^{n-1} & \cos\theta_{n-1}^{n-1} \end{pmatrix} \quad (3.9)$$

If we choose in $E_n$ the spherical coordinates $\theta_1,\theta_2,...\theta_{n-1}$ we write

$$\begin{aligned} \xi_1 &= \sin\theta_{n-1}\ldots\sin\theta_2\sin\theta_1 \\ &\vdots \\ \xi_n &= \cos\theta_{n-1} \end{aligned} \quad (3.10)$$

with $0\le\theta_1\le 2\pi,\ 0\le\theta_j\le\pi,\ j=2,...,n-1$

and $\vec{r}=\vec{r}_n=(x_1,x_2,...,x_n)$, $x_i = r\xi_i$.

We find the Vilenkin parameterization [26] for SO(n) and therefore we shall use the same notations. Any rotation g of the group SO (n) can be set as follows

$$g = g^{(n-1)}...g^{(1)}$$

Where $g^{(k)} = g_1(\theta_1^k)...g_k(\theta_k^k)$ (3.11)

And $g_{(n-1)}(\theta^{(n-1)}_{(n-1)}) = B^0_n$ is the transformation

$$\begin{aligned} x'_{n-1} &= x_{n-1}\cos\theta^{n-1}_{n-1} + x_n \sin\theta^{n-1}_{n-1} \\ x'_n &= x_{n-1}\sin\theta^{n-1}_{n-1} + x_n\cos\theta^{n-1}_{n-1} \end{aligned} \qquad (3.12)$$

**3.4 Parameterization of SU(n)**

In the case of m = 1, the matrix $B^1_n$ is function of two variables and $\det(B^1_n) = 1$. The expression $[B^1_n, A^1_{n-2}] = 0$ means that $B^1_n$ leave invariant $A^1_{n-2}$ and the solution is not unique for n>2. If we parameterize like above the last column by the spherical coordinates $z_i = re^{i\psi_i}\xi_i$, $r = 1$, we write

$$u^k_1 = \begin{pmatrix} e^{-i\psi^k_1} & 0 \\ 0 & e^{+i\psi^k_1} \end{pmatrix},\ u^k_i(\theta^k_i, \psi^k_i) = B^2_n = g_i(\theta^k_i)d_i(\psi^k_i) \qquad (3.13)$$

$$d_i(\psi^k_i) = \begin{pmatrix} I_{n-2} & 0 & 0 \\ 0 & e^{-i\psi^k_i} & 0 \\ 0 & 0 & e^{+i\psi^k_i} \end{pmatrix}$$

$$u = u^{(n-1)}...u^{(1)}$$

Where $$u^{(k)} = u^k_0(\psi^k_1)u_1(\theta^k_1, \psi^k_2)...u_k(\theta^k_k, \psi^k_{k+1}) \qquad (3.14)$$

We can also consider other useful options (22), for example

$$u^k_1 = \begin{pmatrix} e^{-i\psi^k_1} & 0 \\ 0 & e^{+i\psi^k_1} \end{pmatrix},\ u^k_i(\theta^k_i, \psi^k_i) = B^1_n = g_i(\theta^k_i)d_i(\psi^k_i)$$

$$d_i(\psi^k_i) = \begin{pmatrix} e^{-i\psi^k_i} I_{n-1} & 0 \\ 0 & e^{-i(n-1)\psi^k_i} \end{pmatrix}$$

$$u = u^{(n-1)}...u^{(1)}$$

Where $$u^{(k)} = u^k_0(\psi^k_1)u_1(\theta^k_1, \psi^k_2)...u_k(\theta^k_k, \psi^k_{k+1}) \qquad (3.15)$$

$$SU(2) \qquad U_2(a) = A^2_2 = \begin{pmatrix} a_1 & a_2 \\ -\bar{a}_2 & \bar{a}_1 \end{pmatrix} \qquad (3.16)$$

**3.5 Parameterization of SO(6)**

It's known that the Lie algebra of SO(6) and the Lie algebra SU(4) are isomorphic. Therefore, there are a non-singular mapping between the generators of SO(6) and SU(4). Since such mapping preserves the Lie bracket structure, we can deduce a parameterization of SO(3), SO(4**),** SO(5) and SO(6) using the expressions of the generators (2.5) and the harmonic oscillator basis [27-30].

**3.6 The invariant measure on the group SU (n)**

the invariant measure is the result of the product of invariants measure on the sphere $S^{2n-1}$ with n = 1... n. We determine first the invariant measure of the group SO(n) and then for the group SU (n).

### 3.6.1 The invariant measure of the group SO (n) and Euclidean measure

the metric on the sphere $S^n$ is: $ds^2 = dr^2 + r^2[\prod_{i=1}^n d\xi_i^2]$

By use the polar coordinates

$$ds^2 = dr^2 + r^2 d\theta_{n-1}^2 + r^2 \sin^2\theta_{n-1} d\theta_{n-2}^2 + ..... + r^2 \sin^2\theta_{n-1} .. \sin^2\theta_2 d\theta_1^2$$

Hence $\quad dV_s = r^{(n-1)} d\xi = Ar^{(n-1)} \sin^{n-2}\theta_{n-1} ... \sin\theta_2 d\theta_1 ... d\theta_{n-1}$

We choose the constant A so that the measure on the sphere is equal to one.

Since $\quad \int_{-\propto}^{\propto} e^{-x^2} dx = 1/(\pi)^{\frac{1}{2}}$ ,

And in the Cartesian n-dimensional harmonic oscillator we have

$$\int e^{-r^2} \prod_{i=1}^n dx_i = \int e^{-r^2} d\vec{r} = \int_0^{\propto} e^{-r^2} r^{n-1} dr d\xi = \frac{1}{2}(\Gamma(\frac{n}{2})/(\pi)^{\frac{n}{2}}) \int_0^{\propto} d\xi = A$$

So $\quad A = \Gamma(n/2)/2\pi^{n/2}$

Finally $\quad dV_s = A \sin^{n-2}\theta_{n-1} ... \sin\theta_2 d\theta_1 ... d\theta_{n-1} \quad (3.17)$

The invariant measure on the group SO(n) is:

$$dg = A_n \prod_{k=1}^{n-1} \prod_{j=1}^{k} \sin^{j-1} \theta_j^k d\theta_j^k \quad (3.18)$$

With $\quad A_n = \prod_{k=1}^n (\Gamma(k/2)/(2\pi^{k/2}))$

The invariant measure $d\xi$ of SO (n) is the angular part of the measure of product of Cartesian harmonic oscillator.

$$\prod_{i=1}^n e^{-(r_i)^2} d\vec{r}^{\,i} = (\prod_{i=1}^n e^{-(r_i)^2} (r_i)^{i-1} dr_i) d\xi, \; d\xi = (\prod_{i=1}^n d\xi^i)$$

The number of parameters of SO(n) is $n(n-1)/2 = 2^0 (\prod_{i=1}^n i) - n$ with n is the number of parameters $r_i$.

### 3.6.2 The invariant measure of the group SU(n) and Fock-Bargmann spaces

By use of the polar coordinates $z = (z_1, z_2, .. z_n), \; z_i = r\varsigma_i = re^{-i\psi_i}\xi_i$

The metric on the sphere $S^{2n-1}$ is:

$$ds^2 = dr^2 + r^2[\sum_{i=1}^n d(\varsigma_i) d(\bar{\varsigma}_i)]$$

$$ds^2 = dr^2 + r^2[\sum_{i=1}^n d\psi_i^{\,2} \xi_i^{\,2} + d\xi_i^2]$$

Therefore

$$dV_s = d\bar{z} = A(\prod_{i=1}^n \xi_i) \sin^{n-2}\theta_{n-1} ... \sin\theta_2 d\theta_1 ... d\theta_{n-1} d\psi_1 ... d\psi_n \quad (3.19)$$

$$= A(\prod_{i=1}^n \xi_i)(d\xi) \prod_{i=1}^n d\psi_i$$

We deduce the connection between the 2n-dimensional cylindrical basis of harmonic oscillator, the measure of integration of Bargmann spaces of dimension 2n and the measure on the sphere $S^{2n-1}$.

We obtain

$$e^{-r^2}\prod_{i=1}^{n} dz_i d\bar{z}_i = e^{-r^2} r^{2n-1}\prod_{i=1}^{n}\xi_i d\xi_i d\psi_i$$

And therefore we note for the following of this work

$$d\mu(U_n) = \prod_{i=2}^{n} d\mu(z^i)$$

We determine A by observing that

$$\int d\mu(U_n) = \int_0^{\infty} e^{-r^2} r^{2n-1} dr d(\hat{U}_n) = \frac{1}{2}\Gamma(n)\frac{1}{\pi^n}\int d(\hat{U}_n) = A$$

$$A = (\Gamma(n)/2\pi^n).$$

Finally:

$$dV_s = (\Gamma(n)/2\pi^n)\prod_{i=1}^{n}\xi_i \sin^{n-2}\theta_{n-1}...\sin\theta_2 d\theta_1...d\theta_{n-1}d\psi_1...d\psi_n \qquad (3.20)$$

the same arguments for the derivation of the measure of integration of SO (n) remain valid in the case of SU (n), [26-27]. It follows that the measure of integration of the group SU (n) must be taken as the angular part of the measure of product of basis of the cylindrical harmonic oscillators.

$$d\mu(U_n^g) = \prod_{i=2}^{n} e^{-(r_i)^2} dz^i d\bar{z}^i = (\prod_{i=2}^{n} e^{-(r_i)^2}(r_i)^{2i-1} dr_i d(\hat{U}_n^g), \qquad (3.21)$$

With $z^i = (z_1^i, z_2^i, ..., z_i^i), (r_i)^2 = z^i \bar{z}^i$

And $$d(\hat{U}_n^g) = (\prod_{i=2}^{n} d(\hat{U}_i) \qquad (3.22)$$

The number of parameters SU (n) is $n^2 - 1 = 2^1(\sum_{i=2}^{n} i) - (n-1)$ with (n-1) is the number of parameters $\{r_i\}$ and the sum is the dimension of the space. We obtain then the relationship between the measure of Fock- Bargmann space and the measure on the unitary group.

This property is very useful for the calculation of the isoscalar factors of unitary groups, paragraph (9), using the Bargmann spaces, after the introduction of the additional parameters $\{r_i\}$.

## 4- Generating function of the basis SU(2) ⊂SU(3)

In this section we review the Schwinger's method of SU(2) as a useful way for the construction of the generating function of SU(3) and for the remainder of this work.

### 4.1 Schwinger's generating function of 3j symbols of SU(2)

The basis of the representation of the group SU (2) in the space of Fock-Bargmann is

$$\varphi_{jm}(z) = \frac{z_1^{j+m} z_2^{j-m}}{\sqrt{(j+m)!(j-m)!}}, \text{ And } (\varphi_{jm}(z))_c = (-1)^{j-m}\varphi_{j-m}(z) \qquad (4.1)$$

In the product space $\prod_{i=1}^{3}\varphi_{j_i m_i}(z^i)$ the invariants of SU (2) or the elementary scalars are:

$$[z^1 z^2],\ [z^1 z^3],\ [z^2 z^3] \qquad (4.2)$$

With $[xy] = x_1 y_2 - x_2 y_1 \; and \; (xy) = x_1 y_1 + x_2 y_2$

Van der Waerden has determined the invariants of SU(2)

$$H_{(\rho)} = \frac{1}{\sqrt{(J+1)!}} \frac{[z^1z^2]^{j_1+j_2-j_3}[z^1z^3]^{j_1-j_2+j_3}[z^2z^3]^{-j_1+j_2+j_3}}{(j_1+j_2-j_3)!(j_1+j_2-j_3)!(j_1+j_2-j_3)!}$$

$$J_3 = j_1+j_2-j_3, J_2 = j_1-j_2+j_3, J_1 = -j_1+j_2+j_3, \quad (4.3)$$

$$J = +j_1+j_2+j_3.$$

To deduce the 3j symbols of SU (2), we multiply this function by

$$\Phi_{j_1j_2j_3}(t,\tau) = [(J+1)!]^{\frac{1}{2}} t^{J+1} \frac{\tau_1^{J-2j_1}\tau_2^{J-2j_2}\tau_3^{J-2j_3}}{(J-2j_1)!(J-2j_2)!(J-2j_3)!} \quad (4.4)$$

And so we get the generating function of polynomials invariant of SU (2).

$$\text{t}\exp[\text{t}\{\tau_3[z^1z^2]+\tau_2[z^1z^3]+\tau_1[z^2z^3]\}] = \sum_j \Phi_{j_1j_2j_3}(t,\tau)H_{(\rho)}(z)$$

To determine the generating function of the coupling of two angular momentums we replace $z_3^1$ by $-z_3^2$ and $z_3^2$ by $z_3^1$ in the above expression. We obtain the Schwinger's formula

$$G(t,\tau,z) = t\exp[t\{\tau_3[x,y]+\tau_1(zx)+\tau_2(zy)\}] =$$

$$\exp[t\{\tau_3[\frac{\partial}{\partial u},\frac{\partial}{\partial v}]+\tau_1\left(z\frac{\partial}{\partial u}\right)+\tau_2\left(z\frac{\partial}{\partial v}\right)]\}\exp[(ux)+(vy)] \quad (4.5)$$

$$= L(t,\tau,z,(x,y))\exp[(ux)+(vy)$$

This formula can be applied to the calculation of several coupling of angular momentum where the great interest of this method. The introduction of the parameter t in (4.4) is useful for the determination of normalization of the basis in the case of several coupling in angular momentum.

### 4.2 Generating function of the basis SU(2)⊂SU(3)

We make a simple revision of our work [22], which is the extension of Schwinger's construction to the generating function of the basis SU(2)⊂ SU(3). We are coupling three basis of two dimensions harmonic oscillator or two dimensions of Fock-Bargmann space.

We put:

$$z^1 = (z_1^1, z_2^1, z_3^1), \qquad z^2 = (z_1^2, z_2^2, z_3^2)$$

$$z_1 = (z_1^1, z_1^2),\ z_2 = (z_2^1, z_2^2),\ z_3 = (z_3^1, z_3^2) \quad (4.6)$$

With $z_i^j \in C$

The vectors $V_{(t,t_0,y)}^{\lambda\mu}(z^1,z^2)$ of the space $D_{[\lambda,\mu]}$ are eigenfunctions of the Casimir operator of the second order $\vec{T}^2$, the projection of $\vec{T}$, $T_0$ on the z-axis and the hypercharge Y. The eigenvalues of these operators are respectively t (t + 1), $t_0$ and y.

$$T_+ = z_1\frac{\partial}{\partial z_2}, \qquad T_- = z_2\frac{\partial}{\partial z_1},$$

$$T_0 = \frac{1}{2}(z_1\frac{\partial}{\partial z_1} - z_2\frac{\partial}{\partial z_2}),$$

$$Y = z_1 \frac{\partial}{\partial z_1} + z_2 \frac{\partial}{\partial z_2} - 2 z_3 \frac{\partial}{\partial z_3} \tag{4.7}$$

The Casimir operator of second order is:

$$\vec{T}^2 = T_0 (T_0 + 1) + T_+ T_-$$

According to the relation (3.7) we have

$$T_{12} V_{(\alpha)}^{\lambda\mu}(z^1, z^2) = 0 \text{ And } T_{ij} = \sum_k z_k^i \frac{\partial}{\partial z_k^j} \tag{4.8}$$

In what follows we work out two successive coupling to obtain the generating function of the basis of the group SU (2) ⊂SU (3).

The vectors $V_{(\alpha)}^{\lambda\mu}(z^1, z^2)$ belong to the space $D_{t_1} \otimes D_{t_2} \otimes D_{t_3}$ which has the basic elements $\varphi_{j_1 m_1}(z^1)\varphi_{j_2 m_2}(z^2)\varphi_{j_3 m_3}(z_3)$ and has the generating function

$$\exp[\sum_{i=1}^{3} (x^j z^j)] \quad \text{with} (x^j z^j) = \sum_k x_k^j z_k^j .$$

The generating functions of the eigenfunctions $W_{t_0}^t(z_1, z_2)$ of $\vec{T}^2$ and $T_0$ can be Deduced by applying the operator $L_1 = L(t, \tau, Z, (x^1, x^2))$ to $\exp[\sum_{i=1}^n (x^i z^i)]$

We get the first coupling

$$\Psi = t_1 \exp\{t_1[t_2 \Delta_3^{(1,2)}]\} + \tau_1 (Z z^1) + \tau_2 (Z z^2)]\} \exp[(x_3^1 z_3^1 + x_3^2 z_3^2)] \tag{4.9}$$

In this expression we have $\tau = (t_2, \tau_1, \tau_2)$.

We note in the following the minors by $\vec{\Delta}^{(1,2)}$

$$\vec{\Delta}^{(1,2)} = (\Delta_1^{(1,2)} \Delta_2^{(1,2)} \Delta_3^{(1,2)}) = \begin{vmatrix} i & j & k \\ z_1^1 & z_2^1 & z_3^1 \\ z_1^2 & z_2^2 & z_3^2 \end{vmatrix} \tag{4.10}$$

$z_1^i, z_2^i$ And $z_3^i$ by $\Delta_1^i$, $\Delta_2^i$ and $\Delta_3^i$, i=1, 2.

We obtain the second coupling by the application of the operator $L_2 = L(t', \tau', Z', (\tau, x^3))$ to $\Psi$.

So that the relation (4.8) is satisfied we put $Z'_2 = 0$ in the result of the second coupling then we get the generating function of the basis vectors $V_{(t, t_0, y)}^{\lambda\mu}(z^1, z^2)$.

$$t_1 t_1' \exp[t_1 t_2 \Delta_3^{(1,2)} + t_1' t_2' t_1 [Z_2 \Delta_1^{(1,2)} - Z_1 \Delta_2^{(1,2)}] +$$
$$t_1 t_1' \tau_1' Z'_1 [(Z_1 \Delta_1^1 + Z_2 \Delta_2^1)] + t_1 t_1' \tau_2' Z'_1 \Delta_3^1] =$$
$$\sum_{\substack{pqt \\ \lambda\mu}} (-1)^q \frac{\sqrt{(\mu + p + 1)!(\mu + \lambda - q + 1)!}}{\sqrt{(\lambda + 1)(2t + 1)\lambda!}} t_1^{\mu + p + 1} t_1'^{\lambda + \mu - q + 1} Z'^{\lambda}_1 \times$$

$$\frac{t_2^q t'^{\mu-q}_2 Z_1^{t+t_0} Z_2^{t-t_0} \tau'^p_1 \tau'^{\lambda-p}_2}{\sqrt{q!(\mu-q)!(t+t_0)!(t-t_0)!p!(\lambda-p)!}} \times V^{\lambda\mu}_{(t,t_0,y)}(z^1,z^2)$$

With

$$V^{\lambda\mu}_{(t,t_0,y)}(z^1,z^2) = N(\lambda\mu;\alpha)(-1)^q \times \sum_k \begin{pmatrix} r \\ k \end{pmatrix} \frac{(\mu-q)!\,p!}{(\mu-q-k)![p-(r-k)]!}$$
$$\times (z_1^1)^{p-(r-k)} (z_2^1)^{r-k} (z_3^1)^{\lambda-p} (\Delta_1^{(1,2)})^k (-\Delta_2^{(1,2)})^{\mu-q-k} (\Delta_3^{(1,2)})^q$$

The normalization is

$$N(\lambda\mu;\alpha) = \{\frac{(\lambda+1)!(\mu+p-q+1)!}{p!q!(\mu-q)!(\lambda-p)!(\mu+p+1)!(\lambda+\mu-q+1)!} \times \frac{(2t-r)!}{(2t)!r!}\}^{\frac{1}{2}}$$

And

$$y = -(2\lambda+\mu)+3(p+q), \quad 0 \le p \le \lambda,$$
$$t = \frac{\mu}{2} + \frac{p-q}{2}, \qquad 0 \le q \le \mu, \tag{4.11}$$
$$t_0 = t - r, \qquad r = 0,1,\ldots,2t.$$

It is important to note that the normalization of the vector $V^{\lambda\mu}_{(t,t_0,y)}(z^1,z^2)$ results from the coupling method unlike the previous works [7, 10 and 12].

Finally in the notations of the theory of angular momentum we write the first coupling

$$\left|(j_1 j_2) jm\nu\right\rangle\left| j_3 m_3\right\rangle, \ \nu = j_1 - j_2 \ \text{ and } \ \nu = j_1 - j_2$$

the basis of SU (3) is derived from the second coupling $\left|(j_1 j_2) jm;(jj_3) j'm'(\nu+m_3)\right\rangle$,

We put $m' = j'$.

And $\quad V^{\lambda\mu}_{(t,t_0,y)}(z^1,z^2) = \left\langle z^1 z^2 \right\|(j_1 j_2) jm;(jj_3) j' j'(\nu+m_3)\rangle. \qquad (4.12)$

## 5. The basis of SU(3) and the spherical harmonics

It's well known that the generating function of the spherical harmonics may be written in the form:

$$\frac{(\vec{a}\cdot\vec{r})^l}{2^l l!} = [\frac{4\pi}{2l+1}]^{\frac{1}{2}} \sum_m \varphi_{lm}(z) Y_{lm}(\vec{r}) \tag{5.1}$$

$\vec{a}$ is a vector of length zero, $\vec{a}\cdot\vec{a} = 0$ and its components

$$a_1 = -\xi^2 + \eta^2, \ a_2 = -i(\xi^2 + \eta^2), \ a_3 = 2\xi\eta$$

with $\quad \varphi_{lm}(z) = \dfrac{\xi^{l+m}\eta^{l-m}}{\sqrt{(l+m)!(l-m)!}}, z = (\xi,\eta)$

$\varphi_{lm}(z)$ is a base of Bargmann space.

It's easy to deduce from (5.1) the correspondence between the operators $\vec{J}$ of SU (2) and the operators $\vec{L}$ of SO (3).

With $\vec{J} = z^t(\sigma)(\frac{\partial}{\partial z})$ and $\vec{L} = \vec{r} \times \frac{\partial}{\partial \vec{r}}$ (5.2)

$\vec{\sigma}$ Is Pauli's spin ½.

**5.1 The new basis**

In the Bargmann space we write the coupling function in the form:

$$\left\langle z^1 z^2 \left\| (j_1 j_2) jm; (jj_3) j'm'(\nu + m_3) \right\rangle \right. =$$

$$\sum_{m,\nu} C^{j,j_3,j'}_{\nu,m_3,j'} [C^{j_1,j_2,j}_{m_1,m_2,m} \varphi_{j_1 m_1}(z_1^1, z_1^2) \varphi_{j_2 m_2}(z_2^1, z_2^2)] \varphi_{j_3 m_3}(z_3^1, z_3^2)$$

We replace $\varphi_{jm}$ by $Y_{lm}$ in (3.12), and then we write the coupling function in the new form:

$$V^{\lambda\mu}_{(L,m,y)}(\vec{r}_1 \vec{r}_2 \vec{r}_3) = \left\langle \vec{r}_1 \vec{r}_2 \vec{r}_3 \right\| (L_1 L_2) Lm; (LL_3) L'L'(\nu + m_3) \rangle =$$
$$\sum\nolimits_{m,\nu} C^{L,L_3,L'}_{\nu,m_3,L'} [C^{L_1,L_2,L}_{m_1,m_2,m} Y_{L_1 m_1}(\vec{r}_1) Y_{L_2 m_2}(\vec{r}_2)] Y_{L_3 m_3}(\vec{r}_3) \tag{5.3}$$
$$m = m_1 + m_2$$

This new basis of SU (3) is orthonormal and eigenfunctions of $\vec{L}^2$ and $L_z$. We have also the correspondence

$$T_+, T_-, T_0 ---> L_+, L_-, L_z$$
$$\vec{L} = \vec{r}_1 \times \frac{\partial}{\partial \vec{r}_1} + \vec{r}_2 \times \frac{\partial}{\partial \vec{r}_2}.$$

**5.2 The Wigner's D-matrix of the new basis**

It is well known that the Wigner's D-matrix of SU(2) are the generalization of spherical harmonics and may be taken as the angular part of an orthogonal basis, $D^j_{m,m'}(z,z')$, in the four dimension space [22] with integration measure is the cylindrical one. Considering the $D^j_{m,m'}(z,z')$ instead of spherical harmonics we deduce the Wigner's D-matrix of the basis $V^{\lambda\mu}_{(L,m,y)}(\vec{r}_1 \vec{r}_2 \vec{r}_3)$.

$$V^{\lambda\mu}_{(j_1,m,y_1),(j_2,m',y_2)} = \sum\nolimits_{m,m',q} [C^{j_1,j_3,j'}_{\nu,m_3,j'} [C^{j_1,j_2,j}_{m_1,m_2,m}] \; \{\prod\nolimits_{i=1}^3 D^{j_i}_{m_i,m_i'}(z_i, z_i')\}$$
$$[C^{j_1,j_3,j''}_{q,m'_3,j''} C^{j_1,j_2,j'}_{m'_1,m'_2,m'}]$$
$$m = m_1 + m_2,\; m' = m_1' + m_2'$$

We can write this expression using the elements of the matrix of SO (4)

$$D^{(p,q)}_{(jm,j'm')} = \sum_{m,m'} [C^{j_1,j_2,j}_{m_1,m_2,m}] \; \{\sum\nolimits_{i=1}^2 D^{j_i}_{m_i,m_i'}(z_i, z_i')\} [C^{j_1,j_2,j'}_{m'_1,m'_2,m'}]$$
$$p = j_1 + j_2, q = j_1 - j_2$$

So $$V^{\lambda\mu}_{(j_1,m,y_1),(j_2,m',y_2)} = \sum_{m_3,m'_3,q} [C^{j_1,j_3,j'}_{q,m_3,j'} \; D^{(p,q)}_{(jm,j'm')} \; \{D^{j_i}_{m_i,m_i'}(z_i, z_i')\} [C^{j_1,j_3,j''}_{q,m'_3,j''}]] \tag{5.4}$$

We can write also this expression in a new form in the space of Bargmann:

$$V^{\lambda\mu}_{(j_1,m,y_1),(j_2,m',y_2)} = \int d\mu(z)d\mu(z')[V^{\lambda\mu}_{(t,t_0,y)}(z^1,z^2)\times \prod_{i=1}^{3} D^{j_i}_{m_i,m_i'}(z_i,z_i')V^{\lambda\mu}_{(t,t_0,y)}(z'^1,z'^2)] \tag{5.5}$$

By analogy with the spherical harmonic of three dimensions, the function $V^{\lambda\mu}_{(L,m,y)}(\hat{r}_1\hat{r}_2\vec{r}_3)$ can be interpreted as the angular part of $V^{\lambda\mu}_{(t,t_0,y)}(z^1,z^2)$. It is simple to observe also that the integral of three products of these functions may be expressed in terms of 9j symbol [29].

## 6- Generalization of generating function of SU(2) to SU(n)

We observe that the parameters and their powers in the generating function of the basis of SU(2) are linked to the raising and lowering operators and their powers then we generalized it by an empirical way [24] to SU(n) basis. So we found from this new generating function the states max , min of SU (n) and the generating function of SU (3) basis. This function is practical for the derivation of the invariant polynomials of SU(n) from the Gel'fand basis of unitary group SU(3(n-1)).
In the representation of Bargmann-Moshinsky [6-7] we write

$$\sum_{h_{lm}} A_n\phi_n(h,(x,y))P\begin{bmatrix}[h]_n\\(h)_n\end{bmatrix}(\Delta(z)) = \exp\left[\sum_i \phi^i_n\Delta^i_n(z)\right]$$

$$\phi_n(h,(x,y)) = \prod_{l=2}^{n}\prod_{m=1}^{l}\left[\frac{x(m,l)^{L(m,l)}}{\sqrt{L(m,l)!}}\frac{y(m,l)^{R(m,l)}}{\sqrt{R(m,l)!}}\right] \tag{6.1}$$

With $\quad L^m_n = L(m,n) = h_{m,n} - h_{m,n-1}, \quad R^m_n = R(m,n) = h_{m,n-1} - h_{m+1,n}$ (6.2)

$\{\Delta^n_i(z)\}$ Is all minors $\{\Delta(z)\} = \{\Delta^{12\ldots l}_{i_1\ldots i_l},\ i,j = 1,\ldots,n\}$ constructed from the complex matrix for the selection of rows $1,2,\ldots,l$ and columns $i_1,i_2,\ldots,i_l$.

In this representation the Gel'fand basis will be noted by $\Gamma\begin{pmatrix}[h]_n\\(h)_n\end{pmatrix}(\Delta(z))$.

### 6.1 The binary fundamental representation (B.F.R.) $\Delta^n_i(z)$

A simple representation of the fundamental representation basis is very useful and has a great interest not only for calculations but especially for the invariance, which is connected with the complement of binary numbers.
We associate to each miner a table of n-boxes numbered from 1 to n. We put "one" in the boxes $i_1,i_2,\ldots,i_l$ and zeros elsewhere.

$$\Delta^{12\cdots l}_{i_1\cdots i_l} = \overset{\begin{matrix}1 & & i_1 & & i_l & n\end{matrix}}{\left|\begin{matrix}0 & \cdots & 1 & \cdots & 1 & 0\end{matrix}\right|} \tag{6.3}$$

Table (1)

We refer to this new representation by the binary representation which has ($2^n - 2$) elements. $\Delta_i^n(z)$ Is represented simply by the number "i" written in binary an array of n-boxes.

**6.2 Calculus of coefficients $\varphi_n^i(x,y)$**

The coefficients $\varphi_n^i(x,y)$ may be written as product of parameters $y(\mu,\lambda)$ and $x(\mu,\lambda)$ . We determine the indices of these parameters by using the following rules:

a-We associate to each "one" which appeared after the first zero [see Tables (2-3)] a parameter $y(\mu,\lambda)$ whose index $\lambda$ are the number of boxes and $\mu$ the number of "one" before him, plus one.

b-We associate to each zero after the first "one" [see Tables (2-3)] a parameter $x(\mu,\lambda)$ whose index $\lambda$ is the number of boxes and $\mu$ the number of "one" before him.

5.3 Gel'fand labelling and B.F.R. basis

We deduce from the tables (2-3) the expression of $\{\Delta_{i_1 \ldots i_l}^{12\ldots l}\}$ in Gel'fand labelling.

We put:

| | | |
|---|---|---|
| $SU(2)\,B.F.R$ | $\lvert 0 \quad 1 \rvert$ | $\lvert 1 \quad 0 \rvert$ |
| $Gel'\ fand\ basis$ | $\begin{pmatrix} 1 \quad 0 \\ 0 \end{pmatrix}$ | $\begin{pmatrix} 1 \quad 0 \\ 1 \end{pmatrix}$ |
| $\varphi_n^i(x,y)$ | $y(1,2)$ | $x(1,2)$ |

Table (2)

| 0 | 1 | 2 | 3 | 4 | 5 | 6 |
|---|---|---|---|---|---|---|
| $SU(3)$ $B.F.R$ | $\lvert 1 \ 0 \ 0 \rvert$ | $\lvert 0 \ 1 \ 0 \rvert$ | $\lvert 0 \ 1 \ 0 \rvert$ | $\lvert 1 \ 0 \ 0 \rvert$ | $\lvert 1 \ 0 \ 1 \rvert$ | $\lvert 1 \ 1 \ 0 \rvert$ |
| $Gel'\ fand$ $basis$ | $\begin{pmatrix} 1 \ 0 \ 0 \\ 0 \ 0 \\ 0 \end{pmatrix}$ | $\begin{pmatrix} 1 \ 0 \ 0 \\ 1 \ 0 \\ 0 \end{pmatrix}$ | $\begin{pmatrix} 1 \ 1 \ 0 \\ 1 \ 0 \\ 0 \end{pmatrix}$ | $\begin{pmatrix} 1 \ 0 \ 0 \\ 1 \ 0 \\ 1 \end{pmatrix}$ | $\begin{pmatrix} 1 \ 1 \ 0 \\ 1 \ 0 \\ 1 \end{pmatrix}$ | $\begin{pmatrix} 1 \ 1 \ 0 \\ 1 \ 1 \\ 1 \end{pmatrix}$ |
| $\varphi_n^i(x,y)$ | $y(1,3)$ | $y(1,2)x(1,2)$ | $y(1,2)x(2,3)$ | $x(2,1)x(1,3)$ | $y(2,3)x(1,2)$ | $x(2,3)$ |

Table (3)

Conversely, if we have a Gel'fand labelling of the B.F.R. we can deduce the position of 'one' in the table (1) and as a result the corresponding minor.

From a practical point of view we can determine all the elements of the B.F.R. of SU(n) by filling tables of n-boxes by the number 1.2… ($2^n - 2$) expressed in binary.

The compatible elementary scalar invariants which are the determinants of order n

**6.3 Invariance by complementary of binary numbers (R-reflexion)**

We know that each binary number has a complement then we deduce that $\Delta_i^n$ has a complement $\overline{\Delta_i^n}$. Therefore The B.F.R. is invariant by the transformation $\Delta_i^n$ to $\overline{\Delta_i^n}$.

For SU (2) we have the transformation $\varphi_{jm} \to (-1)^{j+m}\varphi_{j-m}$ taken into account that the complement of $[1 \quad 0]$ is $[0 \quad 1]$ and conversely.

For SU (3) we also deduce the R-Conjugation of Gell-Mann (Resnikoff)

$$V^{\lambda\mu}_{(t,t_0,y)} \to (-1)^{y/2-t_0} V^{\lambda\mu}_{(t,-t_0,-y)} \tag{6.4}$$

The expression of complement $\overline{\phi}^i_n$ may be deduced from $\phi^i_n$ by changing y(m, $\ell$) with x($\ell$-m, $\ell$), and y(m, $\ell$) with x($\ell$-m, $\ell$), and it follows that the expression (6, 1) remains invariant by changing $\Delta^n_i, \phi^i_n$ by $\overline{\Delta}^n_i\ \overline{\phi}^i_n$. We call this property of invariance by reflection or complementarily invariance. We also note that in the basis of U(n) the complement of $[1,1,\cdots,1]$ is $|0\rangle$ in the oscillator basis and 1 in the Fock-Bargmann space.

### 7-The U(3) patterns and the SU(2) coupling quantum numbers

We are going to show that the result deduced by Biedenharn et al. [17] by the "tensor pattern" for relationship between U (3) and a triple of SU(2) angular momentum is deduced naturally with our method and we get the exact coupling.
The generating function of SU (3) in the Bargmann -Moshinsky basis is

$$\sum_{h_{\mu\nu}} A_3\varphi_3(h_{\mu\nu},(y,x)\Gamma_3\begin{pmatrix}[h]_3\\(h)_3\end{pmatrix}(\Delta(z)) = \exp[\Delta^{(12)}_{12}y(2,3) + \tag{7.1}$$

$$(\Delta^{(12)}_{23}x(1,2) + \Delta^{(12)}_{13}y(1,2))x(2,3) + (\Delta^1_1 y(1,2) + \Delta^1_2 x(1,2))y(1,3) + \Delta^1_3 x(1,3)]$$

The comparison of this term with the coupling generating function of SU (3)
We obtain:
$y(1,2) = Z_1$, $x(1,2) = Z_2$, $y(1,3) = \tau'_1$, $x(1,3) = \tau'_2$, $z(2,3) = t'_2$, $y(2,3) = t_2$
In the first coupling we associate parameters with their powers:

$$\begin{aligned} &\tau_1 \to j + j_1 - j_2,\ \tau_2 \to j - j_1 + j_2,\\ &t_2 \to R(2,3) \to h_{22} - h_{33} = -j + j_1 + j_2\\ &Z_1 \to R(1,2) \to h_{11} - h_{22} = j + m,\\ &Z_2 \to L(1,2) \to h_{12} - h_{11} = j - m \end{aligned} \tag{7.2}$$

We deduce that $\quad j = \dfrac{h_{12} - h_{22}}{2}, m = h_{11} - \dfrac{h_{12} + h_{22}}{2}$

In the second coupling we associate also to the parameters of their powers

$$\begin{aligned} &\tau'1 \to R(1,3) = h_{12} - h_{23} = -j_3 + j' + j,\\ &\tau'_2 \to L(1,3) = h_{13} - h_{12} = +j_3 + j' - j,\\ &t'_2 \to L(2,3) = h_{23} - h_{22} = +j_3 - j' + j \end{aligned} \tag{7.3}$$

For $Z'_2 = 0$ we obtain:

$$m' = j',\ j' = \frac{(h_{13} - h_{23})}{2},\ j_3 = \frac{h_{13} - h_{12} + h_{23} - h_{22}}{2} \tag{7.4}$$

We note by $\hat{j}_i$ the quantum numbers assumed by Biedenharn and al. [17] for the coupling. We find that $\hat{j}_1 = j_3$, $\hat{j}_2 = j'$, $\hat{j} = j$. The right coupling is $|(jj_3)j'\rangle$ and not $|(j'j_3)j\rangle$ who is supposed by these authors. So the order is not the same and our method clarifies the relationship between the coupling coefficients of the group SU (2) and the representation of SU (3).

## 8-The representation of the special invariants polynomials of SU(n) by the Gel'fand basis SU(3(n-1))

In this work we limit ourselves to the invariant polynomials that are elements of the basis of the representation of the group SU(3 (n-1)). These invariant polynomials are function of what we call the elementary scalars and results from these some of conditions on the indices of the Gel'fand basis.

### 8.1- The invariants polynomials

The invariants polynomials are function of the compatible elementary scalar invariants which are the determinants of order n. taking into account expressions (2.10) and $h_{3(n-1),3(n-1)} = 0$ we deduce that the invariant polynomials have the form:

$$(\rho) = \begin{pmatrix} h_{1n} & & h_{1n} & h_{1n} & 0 & 0 & 0 & 0 \\ & h_{1n} \cdots & h_{1n} & . & & 0 & 0 & 0 \\ & \ddots & \cdot^{\cdot^{\cdot}} & . & . & 0 & 0 & \\ & & h_{1n} & . & . & & & \\ & & & h_{1n} & ..h_{nn} & & & \\ & & & . & . & . & & \\ & & & . & . & & & \\ & & & h_{11} & & & & \end{pmatrix} \tag{8.1}$$

This shape is similar to double pattern of Biedenharn et al. (14).

### 8.2-The generating function of the coupling function

The generating function of the coupling coefficients of SU(n) is given by

$$\sum_{h_{lm}} {}^{s}A_{3(n-1)}(h)^{s}\phi_{3(n-1)}(h,(y,x))P\begin{bmatrix} [h]_{3(n-1)} \\ (h)_{3(n-1)} \end{bmatrix}(\Delta(z)) = \exp\left[\sum_{i} {}^{s}\phi_{n}^{i}\ {}^{s}\Delta_{n}^{i}(z)\right] \tag{8.2}$$

We will determine the first the compatible elementary scalar and the corresponding ${}^{s}\phi_{n}^{i}$.

#### 8.2.1 Identification of elementary scalar ${}^{s}\Delta_{n}^{i}(z)$ and ${}^{s}\phi_{n}^{i}$

We determine the elementary scalars ${}^{s}\Delta_{n}^{i}(z)$ which are the basic elements of the Gel'fand basis of the SU (3 (n-1)). These scalars are formed of three rows of tables, Where each row of (n-1) boxes. The first boxes $\alpha_i$ of the row 'i' contain a "one" and zero elsewhere. The $\alpha_i$ satisfy the following requirements

$$0 \le \alpha_i \le n-1, \ \sum_{i=2}^{3} \alpha_i = n \tag{8.3}$$

We find for SU (3) seven scalar elementary compatible, which are represented by the following tables:

$$\vec{\Delta}^{1} \bullet \vec{\Delta}^{34} = \left|1 \quad 0 \quad 1 \quad 1 \quad 0 \quad 0\right|, \ \vec{\Delta}^{12} \bullet \vec{\Delta}^{3} = \left|1 \quad 1 \quad 1 \quad 0 \quad 0 \quad 0\right|,$$

$$\vec{\Delta}^1 \bullet \vec{\Delta}^{56} = \overline{\left|1\quad 0\quad 0\quad 0\quad 1\quad 1\right|}, \vec{\Delta}^{12} \bullet \vec{\Delta}^{5} = \overline{\left|1\quad 1\quad 0\quad 0\quad 1\quad 0\right|},$$

$$\vec{\Delta}^3 \bullet \vec{\Delta}^{56} = \overline{\left|0\quad 0\quad 1\quad 0\quad 1\quad 1\right|}, \vec{\Delta}^{5} \bullet \vec{\Delta}^{34} = \overline{\left|0\quad 0\quad 1\quad 1\quad 1\quad 0\right|},$$

$$\vec{\Delta}^1 \bullet (\vec{\Delta}^3 \times \vec{\Delta}^5) = \overline{\left|1\quad 0\quad 1\quad 0\quad 1\quad 0\right|}$$

**8.2.2 Determination of $^{s}\phi_n^i$**

We determine for each elementary scalar the corresponding $\phi$ using rules a) and b). In the expressions of $\phi$ we observe that some parameters are not included, this means that y(m, ℓ), and z(m, ℓ) are nulls.

We deduce that the Gel'fand pattern is reduced to 7 indices variables:

$$(\rho) = \begin{pmatrix} h_{13} & & h_{13} & & h_{13} & & 0 & & 0 & & 0 \\ & h_{13} & & h_{13} & & h_{24} & & 0 & & 0 & \\ & & h_{13} & & h_{24} & & h_{34} & & 0 & & \\ & & & h_{13} & & h_{23} & & h_{33} & & & \\ & & & & h_{11} & & h_{22} & & & & \\ & & & & & h_{11} & & & & & \end{pmatrix} \tag{8.4}$$

The invariants polynomials are formed from one term or monomials and function of compatible product of elementary invariant scalars. Different monomials are orthogonal vectors and the normalization has already been calculate for SU (3) in the paper [10] or may be deduced from the norm of bosons polynomials.

## 9-Generating function for isoscalar factors of SU(3)

The calculation of the coupling coefficients is carried out using the formula

$$\left(\prod_{i=}^{3}\left\langle\begin{pmatrix}[h^i]_n\\(h_i)_n\end{pmatrix}\right|\right)\left|H_{(\rho)}\right\rangle = \frac{1}{\sqrt{d_{h^3}}}\left\langle\begin{matrix}[h^1][h^2]\\(h^1)(h^2)\end{matrix}\middle|\begin{matrix}[h^3]\\(h^3)\end{matrix}\right\rangle, \tag{9.1}$$

and by replacing the bases and $H_{(\rho)}$ by their respective generating functions.

It is important to note that the calculation becomes too intractable for the determination of coupling coefficients for groups SU (n), n ≥ 3. We propose a way to circumvent this difficulty (in part) using the connection of the measurement of SU (n) and measurement of Fock-Bargmann space.

**9.1 Gaunt integral and Wigner's 3j symbols**

**9.1.1 Gaunt integral**

The integral of the product of three-D or Gaunt integral is given by

$$\int D^{[h^1]n}_{(h'1),(h1)}(U_n)D^{[h^2]n}_{(h'2),(h2)}(U_n)D^{[h^3]n}_{(h'3),(h3)}(U_n)dU_n = \tag{9.2}$$

$$\sum_{(\rho)}\begin{pmatrix}[h^1]_, & [h^2]_, & [h^3]_,\\(h'_1) & (h'_2) & (h'_3)\end{pmatrix}_{(\rho)}\begin{pmatrix}[h^1]_, & [h^2]_, & [h^3]_,\\(h_1) & (h_2) & (h_3)\end{pmatrix}_{(\rho)}$$

In the formula (7) we deduct:

$$\int T^{(1,2,3)}_{(U_n)}\prod_{i=1}^{3}\left|\begin{pmatrix}[h^i]_n\\(h_i)_n\end{pmatrix}\right\rangle dU_n=\sum_{(h'\rho)}H_{(\rho)}\begin{pmatrix}[h^1]_, & [h^2]_, & [h^3]\\(h'_1) & (h'_2) & (h'_3)\end{pmatrix}\begin{pmatrix}[h^1] & [h^2] & [h^3]\\(h^1) & (h^2) & (h^3)\end{pmatrix}_{(\rho)} \quad (9.3)$$

It follows that the symbols of Wigner are

$$\int\left\langle H_{(\rho)}\right|T^{(1,2,3)}_{(U_n)}[\prod_{i=1}^{3}\left|\begin{pmatrix}[h^i]_n\\(h_i)_n\end{pmatrix}\right\rangle]dU_n=\begin{pmatrix}[h^1] & [h^2] & [h^3]\\(h_1) & (h_2) & (h_3)\end{pmatrix}_{(\rho)} \quad (9.4)$$

So from a practical point of view the left is an invariant polynomial and the right may be derived from the product of three generating functions after transformation and integration. The integration works out in Bargmann space using our parameterization of SU (n).

**9.1.2 Wigner's 3j symbols for SU (2)**

A simple calculation shows that

$$\int T^{(1,2,3)}(U_2)\prod_{i=1}^{3}\exp[r_2(x^i z^i)]d\mu(U_2)=$$
$$\exp[[x^1x^2][z^1z^2]+[x^1x^3][z^1z^3]+[x^2x^3][z^2z^3]] \quad (9.5)$$

we deduct from this expression that

$$\int G(\tau,\bar{z})[T^{(1,2,3)}_{(U_2)}\prod_{i=1}^{3}\exp[r_2(x^i z^i)]]d\mu(U_2)=G(\tau,x). \quad (9.6)$$

this new generating function is depending on the parameters (x) and not function of the variables.

**9. 2 Generating functions for isoscalar factors for SU(3)**

We shall present first the general expression, then the method of calculation and finally We give the result of a particular and interesting case.

**9. 2.1 Expression of generating functions**

It's well known that the coupling coefficients of SU (3) are a product of isoscalar factor by 3j symbol of SU (2).Therefore we use the method of generating function and the expressions (9.3), (9.4) and (9.6) for the determination of the generating function for these isoscalar factors.

For the clarity and to simplify the presentation we put

$$\vec{f}=(x_1u_1,\, x_1u_2,\, x_2),$$
$$\vec{\vec{f}}=(y_1u_2,\, y_1u_1,\, y_2) \quad (9.7)$$

and $y(1,3)=x_1,\ x(1,3)=x_2,\ x(2,3)=y_1,\ y(2,3)=y_2$.

The generating function of SU (3) basis may be written as

$$G((x,y,u),\Delta(z))=\exp[\vec{\vec{f}}\cdot\vec{\Delta}^{(12)}+\vec{f}\cdot\vec{\Delta}^{1}]$$
$$=\sum_{h_{\mu\nu}}A_3\varphi_3(h_{\mu\nu},(x,y,u)\Gamma_3\begin{pmatrix}[h]_3\\(h)_3\end{pmatrix}(\Delta(z)) \quad (9.8)$$

To work in the Bargmann space we must replace $u_1,u_2,x_2,y_1,y_2$ by $r_2u_1,r_2u_2$, $r_3x_2,r_3y_1,r_3y_2$ and $\tau$ by $r_3\tau$ but we keep the same notation for x and y.

By using formulas (9.6-7) we write

$$\int G(r_3\tau, r_2\bar{u})[\int\{(T^{(1,2,3)}(U_3)\prod_{i=1}^{3} G_3^i((y^i, x^i, r_2u), \Delta(z))\}d\mu(U_2)]d\mu(U_3) =$$

$$\int T^{(1,2,3)}(U_3)[\int\{(G(r_3\tau, r_2\bar{u})\prod_{i=1}^{3} G_3^i((y^i, x^i, r_2u), \Delta(z))\}d\mu(U_2)]d\mu(U_3) = \quad (9.9)$$

$$\sum_h \prod_{i=1}^3 A_3^i \prod_{m=1}^2 [\frac{x^i(m,3)^{L^i(m,3)} y^i(m,3)^{R^i(m,3)}}{\sqrt{L^i(m,3)!R^i(m,3)!}}]\times$$

$$\begin{pmatrix} [h^1]_3 & [h^2]_3 & [h^3]_3 \\ (h^1)_3 & (h^2)_3 & (h^3)_3 \end{pmatrix} (\Gamma(\frac{2T+\mu+6}{2})/2)\Phi_{t_1t_2t_3}(1,\tau) \quad (9.10)$$

With $T = t_1 + t_2 + t_3,\ \mu = \mu_1 + \mu_2 + \mu_3$

**9. 2.2 the condition of integration**

To calculate the integration in Bargmann space we apply the formula

$$\Gamma(p) = 2\int_0^\infty \exp(-r_3^2) r_3^{(2p-1)} dr_3$$

p is an integer number.

In the case of the group SU (2), $\mathrm{p} = \mathrm{t}_1 + \mathrm{t}_2 + \mathrm{t}_3$ is an integer.

But in the case of SU (3) the power $r_3$ is

$$2\mathrm{p} - 1 = \sum_{i=1}^3 [2(t_\mathrm{i}) + \mu_i + \lambda_i + q_i - p_i] + 5$$

And $$t_i = \frac{\mu_i}{2} + \frac{p_i - q_i}{2}$$

It follows that $2p = [\sum_{i=1}^3 (\lambda_i + 2\mu_i)] + 6$ must be an integer number.

So it is necessary that $(\sum_{i=1}^3 \lambda_i)/2$ is an integer number therefore this implies that $\sum_{i=1}^3 \lambda_i$ must be an even number.

We have a new selection rule for the isoscalar factors of SU (3).

**9. 3 Generating functions for general case**

The integration of the part between brackets of (9.10) gives generating function invariants by SU(2). Then we classify the results into three classes: the first one is a function of the last column of $U_3$; the second is a function of $\overline{U}_3$ and the third terms Is function of the invariants.

With $(\Delta_3^i)' = \sum_{j=1}^3 u_3^j \Delta^i_j,\ r_3 u_3^j = z_j \quad (\Delta_3^{ij})' = \sum_{j=1}^3 \bar{u}_3^j \Delta^{ij}_j,\ r_3\bar{u}_3^j = \bar{z}_j \quad (9.11)$

And $([\vec{\Delta}^{ij} \times \vec{\Delta}^{kl}]_3)' = \sum_{j=1}^3 u_3^j [\vec{\Delta}^{ij} \times \vec{\Delta}^{kl}]_j \quad (9.12)$

We have also the following formulas:

$$\int \exp[tr_3{}^2]\exp[ar_3\bar{t}]d\mu(t) = \exp[ar_3{}^3]$$

And $$\int \exp[t_i r_3]\exp[a\bar{t}_i]d\mu(t) = \exp[ar_3], i = 1,2,3.$$

Using the above formulas we find the expression

$$\int \exp\{r_3{}^2[(\bar{t}_1 + x_2^1)(\Delta_3^1)' + (\bar{t}_2 + x_2^2)(\Delta_3^3)' + (\bar{t}_3 + x_2^3)(\Delta_3^5)' \quad (9.13)$$

$$+(\tau_1 y_1^1 y_1^2 t[\vec{\Delta}^{12}\times\vec{\Delta}^{34}]'_3+\tau_2 y_1^1 y_1^3 t[\vec{\Delta}^{12}\times\vec{\Delta}^{56}]'_3+\tau_3 y_1^3 y_1^2 t[\Delta^{34}\times\vec{\Delta}^{56}]'_3)]+ \quad (9.14)$$

$$r_3[(y_2^1-\tau_1 y_1^1 x_1^1 t_2-\tau_2 y_1^1 x_1^1 t_3)(\Delta_3^{12})'+(y_2^2-\tau_1 y_1^2 x_1^1 t_1-\tau_3 y_1^2 x_1^3 t_3)(\Delta_3^{34})' \quad (9.15)$$

$$+(y_2^3-\tau_2 y_1^3 x_1^1 t_1-\tau_3 y_1^3 x_1^2 t_2)(\Delta_3^{56})' \quad (9.16)$$

$$+(\tau_1 x_1^1 x_1^2)[\vec{\Delta}^1\times\vec{\Delta}^3]'_3+(\tau_2 x_1^1 x_1^3)[\vec{\Delta}^1\times\vec{\Delta}^5]'_3+(\tau_3 x_1^3 x_1^2)[\vec{\Delta}^3\times\vec{\Delta}^5]'_3]\ + \quad (9.17)$$

$$[x_1^1 y_1^3[\tau_2\vec{\Delta}^1.\vec{\Delta}^{56}+\tau_1\vec{\Delta}^3.\vec{\Delta}^{56}+\tau_3\vec{\Delta}^1.\vec{\Delta}^{34}]+\bar{t}](r_3)^2\}\ d\mu(U_3) \quad (9.18)$$

1-The expressions (9.14-15) expressed in term of $U_3$.

2 - The expressions (9.16-18) expressed as a function of conjugate of $U_3$ or $\overline{U_3}$ .

3 - (9.19) is the invariant by $U_3$.

We do the integrations in Fock-Bargmann space using (2.5) and the formula:

$$(\vec{a}\times\vec{b})\cdot(\vec{c}\times\vec{d})=\begin{pmatrix}\vec{a}\cdot\vec{c} & \vec{b}\cdot\vec{c}\\ \vec{a}\cdot\vec{d} & \vec{b}\cdot\vec{d}\end{pmatrix} \quad (9.19)$$

So we get an expression, function of the elementary scalars and may be deduced by an intuitive way in the case of the group SU (3).

For the final calculation we must use the Bargmann integral

$$\int d\mu(z)\exp(-\bar{z}^t Xz+A^t z+\bar{z}^t B)=(\det X)^{-1}\exp(A^t X^{-1}B) \quad (9.20)$$

Which is valid whenever the hermitian part of X is positive definite. Here z is the Components of the vector $z=(z_1,z_2,\ldots,z_n)$ .

**9. 4 Generating functions for the case $x_1=y_1=0$ .**

We find the following generating function for the special $x_1=y_1=0$

$$\exp\ \{[x_2^2\vec{\Delta}^3\cdot\vec{\Delta}^{56}+x_2^1\vec{\Delta}^1\cdot\vec{\Delta}^{56}][-x_1^1 y_2^2 y_1^3[\tau_2\vec{\Delta}^1\cdot\vec{\Delta}^{56}+\tau_1\vec{\Delta}^3\cdot\vec{\Delta}^{56}]]+$$
$$y_2^3\,[1-y_1^3\tau_2 x_1^1\vec{\Delta}^1\cdot\vec{\Delta}^{56}-y_1^3\tau_1 x_1^2\vec{\Delta}^3\cdot\vec{\Delta}^{56}]]$$
$$\{x_2^1\tau_1 x_1^3 x_1^2\}-x_2^2\tau_2 x_1^1 x_1^3+x_2^3\tau_3 x_1^1 x_1^2\}\vec{\Delta}^1.(\vec{\Delta}^3\times\vec{\Delta}^5)$$
$$+(x_2^3(y_2^2\vec{\Delta}^5\cdot\vec{\Delta}^{34}+y_2^1\vec{\Delta}^5\cdot\vec{\Delta}^{12}+x_2^2 y_2^1\vec{\Delta}^3\cdot\vec{\Delta}^{12}+x_2^1 y_2^2\vec{\Delta}^1\cdot\vec{\Delta}^{34})\}/ \quad (9.21)$$
$$(1-\tau_1 y_1^3 x_1^2\vec{\Delta}^3\cdot\vec{\Delta}^{56}-\tau_2 x_1^1 y_1^3\vec{\Delta}^1\cdot\vec{\Delta}^{56})$$

The development of (9.21) and comparison with (9.11) gives the expression of isoscalaires factors. It is important to emphasize that we can obtain compact useful expressions for special cases of isoscalar factors from the formula (9.1) that we will be published in another paper.

## 10. The Wigner's D-matrix elements for SU(3)

The expression of Wigner's D-matrix elements for SU (3) was given in the appendix of our previous work [22].But we will give more the method of calculation.

**10.1 Generating function for the Wigner's D-matrix elements for SU(3)**

We shall derive the generating function of the Wigner's D-matrix elements for SU(3). We have

$$\langle\Psi'|U_3|\Psi\rangle=\int\exp[\vec{\vec{f}}'\cdot\vec{\Delta}^{(12)}+\vec{f}'\cdot\vec{\Delta}^1]U_3\exp[\vec{\vec{f}}\cdot\vec{\Delta}^{(12)}+\vec{f}\cdot\vec{\Delta}^1]d\mu(z^1)d\mu(z^2) \quad (10.1)$$

Using (3.20) we can write the above expression in the form

$$\langle \Psi' | U_3 | \Psi \rangle = \int \exp[\vec{\lambda}' \cdot \vec{\Delta}^{12} + \vec{\alpha}' \cdot \overline{\Delta}^1] \exp[\vec{\lambda} \cdot \vec{\Delta}^{12} + \vec{\alpha} \cdot \vec{\Delta}] d\mu(z^1) d\mu(z^2)$$

With
$$\begin{pmatrix} \alpha_1 \\ \alpha_2 \\ \alpha_3 \end{pmatrix} = \begin{pmatrix} 1 & 0 & 0 \\ 0 & e^{-i\nu}\cos\frac{\chi}{2} & -\sin\frac{\chi}{2} \\ 0 & \sin\frac{\chi}{2} & e^{+i\nu}\cos\frac{\chi}{2} \end{pmatrix} \begin{pmatrix} x_1 r_1 \\ x_2 r_2 \\ x_2 \end{pmatrix},$$

$$\begin{pmatrix} r_1 \\ r_2 \end{pmatrix} = {}^t U_2(\psi\theta\varphi) \begin{pmatrix} u_1 \\ u_2 \end{pmatrix}, \quad \begin{pmatrix} s_1 \\ s_2 \end{pmatrix} = U_2(\psi'\theta'\varphi') \begin{pmatrix} u_1' \\ u_2' \end{pmatrix},$$

$$\vec{\lambda} = (y_1 s_2, y_1 s_1, y_2).$$

The generating function of D-matrix elements of SU(3) may be calculated by two methods, either by using the Schwinger's method and the formulas (1.4) or by a direct calculation using the Bargmann integral (9.22) .

We have made the calculation by both method but we present the first because it is faster. In applying the formula (1.4) twice on the expression

$$\frac{\partial \langle \Psi' \| \Psi \rangle}{\partial \alpha_i} = \langle \Psi' | \Delta_i^1 | \Psi \rangle = \hat{\Delta}_i^1 \tag{10.2}$$

We find a linear system of $\hat{\Delta}_i^1$, as variables, which the resolution gives the following expressions:

$$\hat{\Delta}_1^1 = \frac{\alpha'_1 - \lambda_1 \vec{\alpha}' \cdot \vec{\lambda}'}{(1 - \vec{\lambda} \cdot \vec{\lambda}')} \langle \Psi' \| \Psi \rangle, \; \hat{\Delta}_2^1 = \frac{\alpha'_2 - \lambda_2 \vec{\alpha}' \cdot \vec{\lambda}'}{(1 - \vec{\lambda} \cdot \vec{\lambda}')} \langle \Psi' \| \Psi \rangle, \; \hat{\Delta}_4^1 = \frac{\alpha'_3 - \lambda_3 \vec{\alpha}' \cdot \vec{\lambda}'}{(1 - \vec{\lambda} \cdot \vec{\lambda}'')} \langle \Psi' \| \Psi \rangle$$

Using (3.18) we find

$$\int \exp[\vec{\lambda}' \cdot \vec{\Delta}^{12} + \vec{\alpha}' \cdot \overline{\Delta}^1] \exp[\vec{\lambda} \cdot \vec{\Delta}^{12} + \vec{\alpha} \cdot \vec{\Delta}] d\mu(z) d\mu(z) =$$

$$\frac{1}{(1 - \vec{\lambda} \cdot \vec{\lambda}')^2} \exp[\frac{\vec{\alpha} \cdot \vec{\alpha}' - (\vec{\alpha} \cdot \vec{\lambda})(\vec{\alpha}' \cdot \vec{\lambda}')}{(1 - \vec{\lambda} \cdot \vec{\lambda}')}] \tag{10.3}$$

The transformation introduced by Prakash et al. [4, 12] does not give this result and our expression is much simpler than that obtained by those authors.

**10.2 Expression of Wigner D-Matrix of SU (3)**

To determine the expression of Wigner D- matrix of SU (3) we are developing the generating function using the following formulas:

$$(1 - x)^{-1-\alpha} = \sum_{p=0}^{\infty} \frac{(\alpha + p)!}{\alpha! \, p!} x^p, \tag{10.4}$$

And
$$\mathrm{U}_2 \varphi_{jm}(u_1, u_2) = \sum_{m'} D^j_{(m',m)}(\Omega) \varphi_{jm}(u_1, u_2), \; (\Omega) = (\psi\theta\varphi) \tag{10.5}$$

Thus we find the expression of the matrix elements of SU (3)

$$\left\langle \begin{matrix} \lambda\mu \\ y't't_0 \end{matrix} \middle| U_3 \middle| \begin{matrix} \lambda\mu \\ ytt_0 \end{matrix} \right\rangle = \frac{1}{NN'} \frac{(-1)\lambda - p'}{p'!(\lambda - p')!(\lambda + 1)!(\mu - q')!} \sum_{ijk} \frac{(\lambda + \mu - q' + l + 1)!}{l!} \times$$

$$\frac{\sqrt{(l)!(\mu-q'-i)!}\sqrt{(p'-j)!(\lambda-p'-j)!}\sqrt{(j+\mu'-q'-i)!(i+p-j)!}\sqrt{(i+j)!(\mu-q-i+p-j)!}}{\sqrt{(j_1+m_2)!(j_1-m_2)!}\sqrt{(j'_2+m'_2)!(j'_2-m'_2)!}}\times$$

$$D^{t'}_{(t'_0,t'_0)}(\Omega')d^{j_1}_{(m_1,m_2)}(\nu_3)d^{j_2}_{(m'_1,m'_2)}(\nu_3)e^{-i\chi(m_2+m'_2)}D^{t}_{(t_0,t_0)}(\Omega)$$

$$2j_1=l+\mu'-q'-i,\ 2m_1=l-\mu'+q'-i,\ 2j_2=-j+\lambda'-k,\ 2m'_1=p'-j-\lambda'+p'+k$$

$$\lambda=\lambda',\quad \mu=\mu',\ k+l=q'$$

$$2t'=\mu'-q'+p',\ 2t=\mu-q+p,\ 2t'_0=\mu'-q'-p'-2(i-j),\ 2t=2(i+j)-\mu+q-p$$

## References

[1] T. Tilma and E.C.G. Sudarshan,"Generalized Euler angle parameterization for SU(N)" J. Phys. A 35 (2002)10467

[2] T. Tilma and E.C.G. Sudarshan, "Usage of an Euler angle parameterization of SU(N) and U(N) for Entanglement Calculations on a Two-Qubit System" J. Phys. Soc. Jpn. 72 (2003) Suppl. C 181

[3] S. Bertini, S.L. Cacciatori and B.L. Cerchial, 'On the Euler angles for SU(N)' arXiv: math-phy/0510075v2 (2005)

[4] J.S. Prakash 'Wigner's D-matrix elements for SU(3)-A generating Function' arXiv: math-phy/9604036 (1996)

[5] D.J. Rowe, B.C. Sanders, H. de Guise arXiv: math-phy/9811012v2 (1996, 2006)

[6] M. Moshinsky, J. Math. Phys. 4(1963)1128;

[7] M. J. Moshinsky, Rev. Mod. Phys. 34(1962)813

[8] T.A. Brody, M. Moshinsky and I. Renero, J. math. Phys. 6(1965)1540

[9] J. Nagel and M. Moshinsky, J. math. Phys. 6(1965)682

[10] M. Resnikoff, J. math. Phys. 8(1967)63

[11] C. K. Chew and R. T. Sharp, Can. J. Phys., 44 (1966) 2789

[12] J. S. Prakash and H. S. Sharatchandra, arXiv: math-phy/960101 v1 (1996)

[13] G.E. Baird and L.C. Biedenharn, J. math. Phys. 4(1963)1449

[14] J.D. Louck Am. J. Phys. 38, 3, 1970

[15] L.C. Biedenharn and Holmann III Group Theory and its Applications (1970), Ed. Loebl (New York: Academic Press)

[16] L. C. Biedenharn, M. Lohe and J.D. Louck, J. math. Phys. 26(1985)1458

[17] L. C. Biedenharn, Chen, J.D. Louck, M. Lohe, 'The Role of SU(2) 3n-j Coefficients in SU(3)',Proceedings of Conference "Symmetry and Structural Properties, "Pozana, Poland, September 1994, in: Los Alamos preprint 1995.

[18] R. Le Blanc and D.J. Rowe, J. Phys. A, 19(1986)1624

[19] R. Le Blanc and K.T. Hecht, J. Phys. A, 20 (1987)4613

[20] D.J. Rowe and C. Bahri, J. Phys. A, 41, 9 (2000)6544

[21] C. Bahri, D.J. Rowe, J. P. Draayer, Computer Physics Communications 159(2004)121

[22] M. Hage-Hassan, J. Phys. A, 12(1979)1633

[23] V. Bargmann, Rev. Mod. Phys. 34(1962)829

[24] M. Hage-Hassan, J. Phys. A, 16 (1983)1835

[25] M. Hage-Hassan, J. Phys. A, 41, 9(1983)2891

[26] N. J. Vilenkin Fonctions spéciales et théorie de la représentation des groupes

Dunod (1991).
[27] A. O. Barut and R. Raczka, Theory of group representations and applications PWN- Warszawa (1980).
[28] J. Schwinger, in quantum Theory of Angular Momentum, Ed. L.C. Biedenharn and H. Van Dam (New York: Academic Press)
[29] L. C. Biedenharn, J. math. Phys. 6(1962)682
[30] M. Hage-Hassan, arXiv: math-phy/0610021 (2006)